\documentclass[showpacs, twocolumn]{revtex4-1}
\usepackage{graphicx}
\usepackage{amsmath}
\usepackage{csquotes}

\begin{document}

\title{Quantum nature of gravity in self-bound quantum droplets}

\author{  Asma Tahar Taiba$^{1}$ and Abdel\^{a}ali Boudjem\^{a}a$^{2,3}$}

\affiliation{$^1$  LPTHIRM, Department of Physics, Faculty of Sciences, University of Blida 1,  P.O. Box. 270, 09000, Blida, Algeria \\
$^2$ Department of Physics, Faculty of Exact Sciences and Informatics, 
and $^3$ Laboratory of Mechanics and Energy, Hassiba Benbouali University of Chlef, P.O. Box 78, 02000, Chlef, Algeria.} 

\email {a.boudjemaa@univ-chlef.dz}

\date{\today}

\begin{abstract}

We explore the possibility of testing the quantum nature of the gravitational field with an ultracold self-bound quantum droplet of one-dimensional Bose-Bose mixtures.
To this end, we solve variationally and numerically the underlying generalized Gross-Pitaevskii equation which includes the effects of quadratic and cubic nonlinearities.
We derive the associated generalized uncertainty principle  and its corresponding minimal length.
The obtained modified uncertainty relation enables us to search for the quantum gravity signatures in both small and large droplets.
We place bounds on the parameter using existing experimental data from recent experiment of dilute droplets of potassium.
Improved upper bounds on the generalized uncertainty principle parameters are found from our analysis.

\end{abstract}

	\maketitle

\section{Introduction}

The search for a quantum gravity (QG) has inspired various theoretical frameworks, including loop quantum gravity, string theory, and doubly special relativity \cite{penrose2014,Amati,Penrose1,Brons}. While each of these theories has significant progress, they continue to face serious obstacles
most notably the absence of a direct experimental testing and the difficulty of recovering classical gravity in the low-energy regime \cite{Bose,Belen,Howl,Piko}. 
Many QG models predict a minimal measurable length which transforms the canonical Heisenberg uncertainty principle (HUP) \cite{Heisenberg}, 
involving position and momentum standard deviations, into the generalized uncertainty principle (GUP) \cite{Ali,Scard1,Mag}. 
This modification implies fundamental changes in the canonical commutation relations \cite{Kempf}.
The need for a QG theory is particularly important at the small scale, around Planck scale ($10^{-35}$ m), where it is expected to solve problems in general relativity, such as the behavior of matter inside black holes \cite{Scar,Casa}. 
Several experimental approaches have been proposed in order to investigate the quantum features of gravity by utilizing fundamental ideas from quantum mechanics (see e.g.\cite{Brack,Scard,Bishop,Gao,Fuchs,Piko,Simon}). 
On the other hand, ultracold neutral atoms or Bose-Einstein condensates (BEC) constitute a fascinating experimental platform for testing and exploring QG at low energies
due to their flexibility in controlling different parameters \cite{Howl, Shir, Bris, Bris1, Dos,Hans,Jaf,Simon,Das, Boudj, Boudj1}.

One of the most exciting developments in the field of ultracold atoms is the recent prediction of quantum droplets \cite{Petrov,Petrov1}.
Such self-bound states  have been realized both in dipolar BECs \cite{ Pfau,Pfau1, Chom}, and in binary BEC mixtures of potassium \cite{Cab, Sem}.
The exploration of quantum droplets with their exquisite properties opens a new avenue for probing quantum nature of gravity.
These exotic quantum composites are stabilized owing to the delicate competition between attractive mean-field interaction (corresponding to a
negative vacuum pressure) and  repulsive beyond mean-field quantum fluctuations  (corresponding to a positive vacuum pressure) originating from the Lee-Huang-Yang (LHY) correction \cite{LHY},
similar to the  emergence of gravitationally bound systems such as photon and neutron stars. 
The equilibrium state of these latter is obtained owing to the quantum photons and neutron pressures in a HUP which halt the gravitational attraction \cite{Bors}. 
The droplet behaves in a completely coherent way enabling the exploration of QG and the formation of spacetime.
Quantum fluctuations are important for studying the early universe hence, providing an ideal setting to test QG.
Furthermore, quantum droplets are self-bound objects and do not need any external potential to remain intact. They are characterized by a constant density and surface effects. 
The droplet has a negative chemical potential to prevent evaporation.
Their ultradilutness make it plausible to bring self-bound droplets into a regime where quantum and gravitational effects are relevant, allowing to quantize gravity.
From these perspectives, self-bound quantum droplets  provide long-lived testbeds with a high degree of isolation and control for 
bridging the gap between quantum mechanics and general relativity.

The aim of this paper is then to explore the possibility of testing the quantum aspects of the gravitational field with an ultracold self-bound quantum droplet of one-dimensional (1D) Bose mixtures.
The idea is to formulate a GUP which adds gravitational effects to the standard HUP in the framework of the generalized Gross-Pitaevskii equation (GGPE) containing
competing cubic and quadratic nonlinearities, and governing the static and the dynamics of self-bound droplets.
Note that the GUP  has been also constructed based on different nonlinear Schr\"odinger equations (see e.g. \cite{Jack,Rud,Bra,Budi}).
In the context of ideal and weakly-interacting BECs, the GPU has been used in order to search for QG signatures (see, e.g., \cite{Fit, VaK,Zhang, Cast, Li, Sanj, Das, Boudj, Boudj1}). 

First, we apply a super-Gaussian variational method to the GGPE, to derive the corresponding GUP and study the emerging bound states. 
We analyze the influence of QG corrections on the binding energy and on the equilibrium size of both small and large droplets.
The results reveal that our theory involves a minimal length notably for large droplets. 
We then delve into the numerical simulation of the GGPE, we show that the QG effects may strongly modify the density profiles especially in the flat-top region
and reduce the droplet width.
Finally, employing recent results from the ${}^{39}$K atom experiment \cite{Sem}, we constrain upper bounds on our GUP deformation parameters
and on the minimal length which may be probed via future advanced techniques.

\section{Quantum droplets with the GUP}

\subsection{Model}

We consider a 1D symmetric Bose-Bose mixture with equal masses, $ m_1=m_2=m$ and equal intraspecies coupling constants $g_1 =g_2 = g$.
Such a symmetry allows the system to be effectively reduced to a single-component description, simplifying the theoretical treatment.
The dynamics of the condensate wavefunction $\psi(x,t)$ is governed by the GGPE that incorporates quantum fluctuation effects via the LHY corrections \cite{Petrov1,Boudj3}:
\begin{equation} \label{GGPE}
i\hbar \partial_t \psi = -\frac{\hbar^2}{2m} \partial_x^2 \psi + \delta g |\psi|^2 \psi - \frac{\sqrt{2m}}{\pi \hbar} g^{3/2} |\psi| \psi,
\end{equation}
where $\delta g = g_{12} + \sqrt{g_{1} g_{2}} > 0$ represents the imbalance between inter- and intra-species interactions, and $g = \sqrt{g_{1} g_{2}}$. 
The competition between the mean-field attraction ($\delta g |\psi|^2 \psi$) and the LHY repulsion ($-g^{3/2} |\psi| \psi$) stabilizes the droplet against collapse \cite{LHY}.

\subsection{ Generalized Uncertainty Principle}

Many QG models predicted the existence of a minimum measurable length $(\Delta x)_{\min}$ \cite{Mag,Das2}. 
This leads to a modification of the standard HUP into the GUP at energies close to the Planck energy
scale $E_P$. Here we consider the so-called Kempf-Mangano-Mann  (KMM) proposal, which was first discussed in Ref. \cite{Mag}:
\begin{equation} \label{GUP}
\Delta x \, \Delta p \geq \frac{\hbar}{2} \left(1 + \beta (\Delta p)^2 \right),
\end{equation}
where $ \beta=\beta_0 \ell_p^2/\hbar^2= \beta_0/(M_P c)^2$ is the  GUP deformation parameter with $M_P = \sqrt{\hbar c/G}$ being the Planck mass, 
 $\ell_p=\sqrt{G \hbar/c^3}$ is the planck length, $G$ being the gravitational constant and $c$ is the speed of light in vacuum.
The minimal observable length results in from Eq.~(\ref{GUP}) reads, $ (\Delta x)_{\text{min}}= \hbar \sqrt{\beta}=\sqrt{\beta_0}\ell_p$.

According to Ref.~\cite{Das2} we can introduce a set of canonical operators $x_{0i}$ and $p_{0i}$, which satisfy a standard commutation relation $[x_{0i}, p_{0j}]= i\hbar \delta_{ij} $.
This implies that
\begin{equation}\label {GUP1}
x_i= x_{0i},   \;\;\;\;\;\;\;  p_i= p_{0i} (1+ \beta p_0^2),
\end{equation}
where $p=\sqrt{p_{0i} p_{0i}}$.

It is woth stressing that the KMM  can be extended to systems made of $N$ identical particles (e.g. Bose gases, superfluids, quantum droplets, $\cdots$ )
with coordinates $x_i$ and momenta $p_i$. Therefore, the center of mass coordinates $X_i$ and the total momentum $P_i$ of the system are defined as: 
$X_i= N^{-1}\sum_{l =1}^N x_i^l$  and  $P_i= \sum_{l =1}^N p_i^l$ \cite{Fadel}.
If the GUP applies to the composants of a rigid macroscopic body where $p_{l} \sim |\mathbf P|/N$ and rescaling $\beta$ as $\beta/N^2$, we get \cite{Fadel}
\begin{equation} \label{GUP11}
\Delta X \, \Delta P \geq \frac{\hbar}{2} \left(1 + \beta (\Delta P)^2 \right).
\end{equation}
which has exactly the same  algebraic structure as the single-particle KMM model (\ref{GUP}).\\
Evidently, this shows that  the GUP may be obtained without assuming modified commutation relations, giving rise to circumvent theoretical difficulties including the soccer ball problem
\cite{Fadel, Matt}.

{\it For our convenience,  we use from now on  the $x$-coordinates instead of $X$-coordinates.}

\subsection{GUP-generalized Gross-Pitaevskii equation}

Within the above redefinitions, the time-dependent GGPE (\ref{GGPE}) turns out to be given as:
\begin{equation} \label{GGPE-GUP}
i\hbar \partial_t \psi = -\frac{\hbar^2}{2m} \partial_x^2 \psi + \beta \frac{\hbar^4}{m} \partial_x^4 \psi + \delta g |\psi|^2 \psi - \frac{\sqrt{2m}}{\pi \hbar} g^{3/2} |\psi| \psi.
\end{equation}
The 1D time-independent GGPE in terms of the new variables $x \rightarrow x/\left(\pi \hbar^2 \sqrt{\delta g}/\sqrt{2m} g^{3/2} \right)$, 
$t \rightarrow t/ \left(\pi^2 \hbar^3 \delta g/2m g^3\right)$ and $\psi \rightarrow \psi /\left(\sqrt{2m}g^{3/2}/\pi \hbar \delta g \right)$, can be written
in the following dimensionless form:
\begin{equation}\label{GGPE-GUP1}
i \partial_t \psi +\frac{1}{2} \partial_x^2 \psi - \bar{\beta} \partial_x^4 \psi - |\psi|^2 \psi + |\psi| \psi=0,
\end{equation}
where $\bar{\beta} = \beta \hbar^2 / \left(\pi \hbar^2 \sqrt{\delta g}/\sqrt{2m} g^{3/2} \right)^2$. \\
The ground-state solution of the 1D GGPE (\ref{GGPE-GUP1}) with $\bar{\beta}=0$ can be written as \cite{Petrov1}:
$$\psi(x) = -\frac{3\mu \exp(-i\mu t)}{1 +\sqrt{ 1+ 9\mu/2} \cosh(\sqrt{-2\mu x^2})},$$
where $\mu$ is the chemical potential.

The energy functional corresponding to Eq.~ (\ref{GGPE-GUP1}) takes the form:
\begin{align}\label{engyfun}
E=\int_{-\infty}^\infty \bigg [\frac{1}{2} \left|\frac{\partial \psi}{\partial x}\right|^2+\bar \beta\left|\frac{\partial \psi}{\partial x}\right|^4 - \frac{2}{3}|\psi|^3+\frac{1}{2}|\psi|^4\bigg].
\end{align}
The minimization of such an energy allows us to calculate the variational parameters and thus provides an estimate of the ground-state energy and the wavefunction of the droplet under the GUP.



\section{ Self-bound quanum droplet-GUP}

\subsection{Variational approximation: super-Gaussian ansatz}

Since the droplet features with a broad flat-top plateau, it is then convenient to adopt a super-Gaussian trial function in order to obtain useful relations between the droplet 
and the GUP parameters. 
It can be written as \cite{Karl,Tsy,Baiz}:
\begin{equation} \label{TDvar}
    \psi = A  \exp{\left[-\left(\frac{x}{2 q}\right)^{2m} + i \alpha x^2 + i \theta \right]},
\end{equation}
where $m$ is the super-Gaussian index, $A$ is the amplitude and $ q$, $\alpha$, and $\theta $ are the width, chirp, and the phase, respectively.
For $m=1$, the ansatz (\ref{TDvar}) reduces to the standard Gaussian trial function capturing the main properties of only small droplets.
The normalization condition, $N=\int_{-\infty}^{+\infty} |\psi|^2 dx$,  implies
 \begin{align}  \label{Norsup}
 	A=\sqrt{ \frac{N} {2q \Gamma (1+s)} }, 
 \end{align}
 where  $\Gamma[z]$ is the gamma fonction, and $s=1/(2m)$.


Substituting the function (\ref{TDvar}) into Eq.~(\ref{engyfun}) and integrating over space, we obtain for  the energy
 \begin{align} \label{TDlagsup}
\frac{E}{N}&=\frac{2^{-s-2}N} {q s \Gamma (s)}+ \frac{\Gamma(2-s)} {8q^2s^2\Gamma(s)} - 2^{s+\frac{1}{2}} 3^{-s-1} \sqrt{\frac{N}{qs \Gamma (s)}}\\
&+  \frac{ q^2\Gamma(3s)} {\Gamma(s)} 2\alpha^2+\frac{\bar\beta N 2^{-5 s-6}} {q^2 s^2 \Gamma(s)^2}\nonumber\\
&\times\bigg(\frac{2^{8 s-3} \Gamma (4-3 s)}{q^3 s^3}+32 \alpha ^2 \left(16 \alpha ^2 q^3 s \Gamma (5 s)+q 4^{2 s}
   \Gamma (s)\right)\bigg) \nonumber.
  \end{align}
Here we used $\Gamma (1+s)=s\Gamma (s) $.

\begin{figure}[htpb]
    \centering 
    \includegraphics[scale= 0.4]{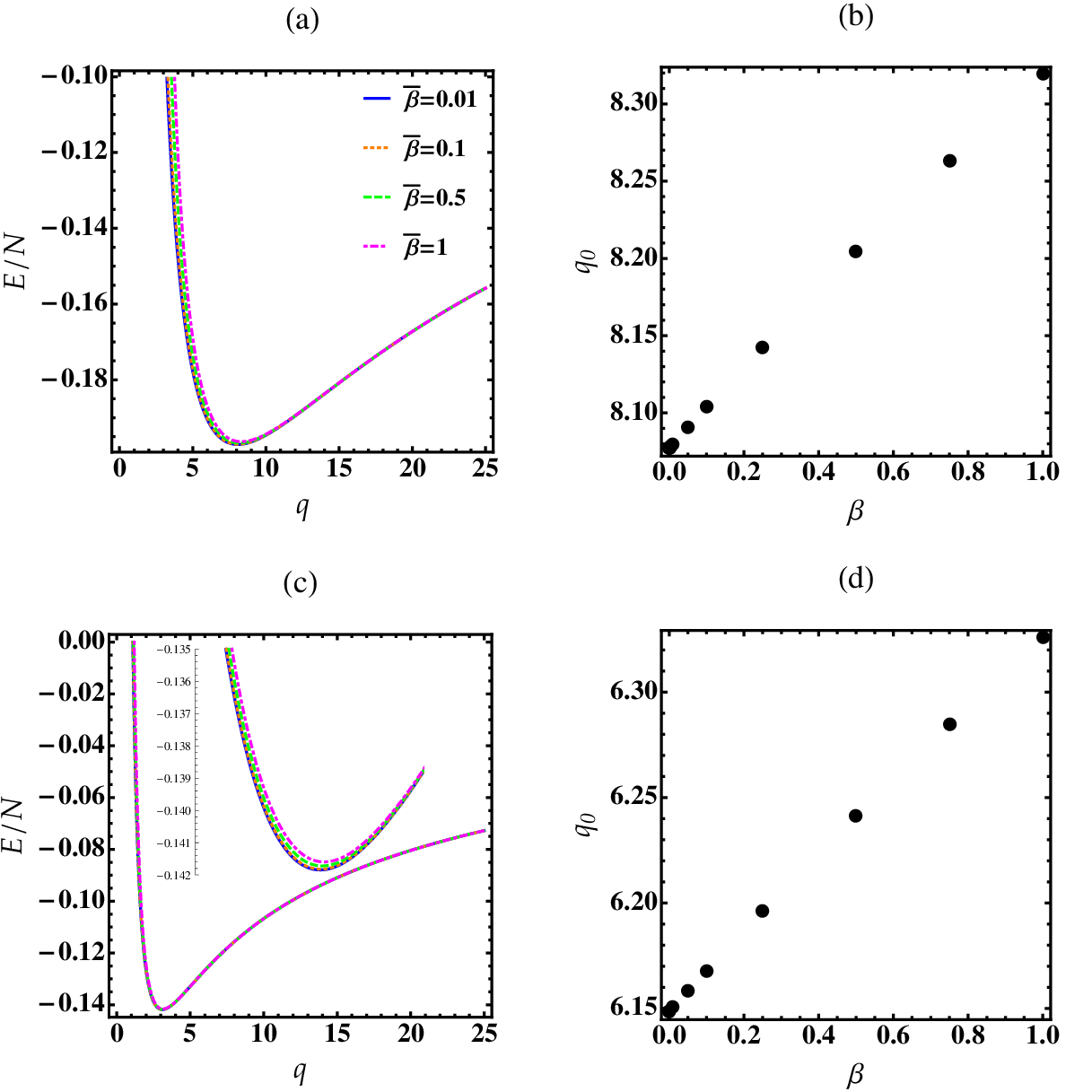}  
\caption{ (a) Energy functional, $E/N$,  of the quantum large droplet versus its width for different values of QG parameter, $\bar\beta$.
(b) Equilibrium width, $q_0$ as a function of  $\bar\beta$.
Parameters are: $N=5$, $m=4$ and $\alpha=0.001$.
(c)-(d) The same as (a) and (b) but for a small droplet with $N=1$, and $m=1$.
Here we used a relatively large value of $\bar\beta$, ($\bar\beta=1$) to highlight the energy shift.}
    \label{fig1}
\end{figure}

The energy (\ref{TDlagsup}) possesses a minimum at a finite value of $q$ corresponding to the equilibrium width $q_0$ of the flat-top droplet (see Fig.\ref{fig1} (a)).
The fixed values of the Gaussian exponent and the norm associated with such stationary solutions are respectively, $m=m_0=4$, and $N=5$. 
We see that the equilibrium width increases monotonically with the GUP parameter, $\bar{\beta}$ (see Fig.\ref{fig1} (b)). 
This indicates that a more stable self-bound droplet requires lower  $\bar{\beta}$.
For small droplets of a Gaussian-like shape with $N=1$ and $m_0=1$, the depth of the local minimum and the equilibrium width 
are less sensitive to the GUP parameter (see Figs.\ref{fig1} (c) and (d)).

\subsection{Super-Gaussian wave-packets and minimal uncertainty}

In this section we show that the GUP model is relevant for static and dynamics of  1D quantum droplets. 
To achieve this, we will derive the generalized uncertainty relation starting from the variational ansatz (\ref{TDvar}). 

Working in Fourier space, Eq.~(\ref{GGPE-GUP1}) takes the form:
\begin{equation} \label{GGPE-GUPK}
i \partial_t \tilde{\psi} = \left (\frac{k^2}{2}  + \bar{\beta} k^4 \right)\tilde{\psi} + F[|\psi|^2 \psi] - F[|\psi|\psi],
\end{equation}
where $F[\psi]=\tilde \psi (k)= (1/\sqrt{2\pi})\int_{-\infty}^{\infty} \psi (x) \, e^{-i k x} \, dx$ is the Fourier transform of $\psi(x)$.
From the linear term of Eq.~(\ref{GGPE-GUPK}), we can define the generalized momentum $K$ as:
$ K^2 = k^2 + 2\bar{\beta} k^4$. For a very small $\bar{\beta}$, we get: 
\begin{equation} \label{Mom}
K \approx k + \bar \beta k^3,
\end{equation}
meaning that the momentum $K$ has limited values.\\
In order to derive the GUP we have to calculate the product $\Delta x \Delta K$, where 
the position and momentum uncertainties are defined as:
\begin{equation} \label{Uncer}
\Delta x = \sqrt{\langle x^2 \rangle - \langle x \rangle^2}, \quad \Delta K = \sqrt{\langle K^2 \rangle - \langle K \rangle^2},
\end{equation}
where the expectation values are given by:
\[
\langle x^2 \rangle = \frac{\int_{-\infty}^{\infty} x^2 |\psi(x)|^2 \, dx}{\int_{-\infty}^{\infty} |\psi(x)|^2 \, dx}, \quad
\langle K^2 \rangle = \frac{\int_{-\infty}^{\infty} K^2 |\tilde{\psi}(K)|^2 \, dK}{\int_{-\infty}^{\infty} |\tilde{\psi}(K)|^2 \, dK}.
\]

\begin{figure}[!h]
    \centering
    \includegraphics[scale = 0.8]{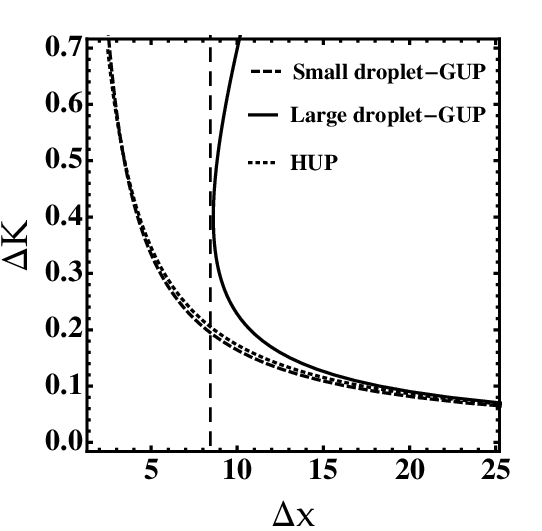}
    \caption{Droplet-GUP from the super-Gaussian wavepackets for $\bar\beta=0.5$ and its corresponding equilibrium width $q=q_0=8.25$ for large droplet and $q=q_0=5.2$ for small droplet.
Dotted line is the HUP for comparison. It is important to note that the obtained large droplet-GUP has no minimal length at $(\Delta x)_{\text{min}}= 8.5$.
However the small droplet-GUP has no minimal length.}
    \label{fig2}
\end{figure}

Now let us apply the obtained results for the super-Gaussian wave-packets (\ref{TDvar}). 
Incorporating  the function  (\ref{TDvar}) and its Fourier transform into Eq.~(\ref{Uncer}), we obtain the uncertainties up to leading order in $\bar\beta$.\\
For a large droplet with $m=4$, we get:
\begin{subequations}\label{Uncer1}
\begin{align} 
&\Delta x =q  \sqrt{ \frac{ \sin \left(\pi /8\right) \Gamma \left(3/4\right)} {\sqrt{\pi }}},\\
&\Delta K \approx \sqrt{ \frac{7 \pi  \csc ^2\left(\frac{\pi }{8}\right)}{64\ 2^{7/8}  \Gamma \left(\frac{9}{8} \right)q}+ \frac{4725 \bar \beta \left(3 \sqrt{2}-2\right) \pi ^{3/2}   \csc \left(\frac{\pi }{8}\right)}{512\ 2^{7/8}  \Gamma \left(\frac{9}{8}\right) \Gamma \left(\frac{7}{4}\right) q^3}}. 
\end{align}
\end{subequations}
Introducing the obtained squared momentum and position terms results the modified GUP:
\begin{equation} \label{GUP2L}
\Delta x \Delta K \approx  \sqrt{0.36 q} \left[1+ \frac{103}{2q} \bar\beta (\Delta K)^2\right],
\end{equation}
which implies that the uncertainty $\Delta x$ is bounded by 
\begin{equation} \label{GUP22L}
(\Delta x)_{\text{min}} \approx \sqrt{ 18.5 \bar\beta}.
\end{equation}
For a small droplet with $m=1$, one has:
\begin{equation} \label{GUP2S}
\Delta x \Delta K \approx   \frac{\sqrt{\pi/2} }{4}q \left[1+ \frac{3} {q\sqrt{\pi/2}} \bar\beta (\Delta K)^2 \right].
\end{equation}
Minimizing $\Delta x$ in Eq.~(\ref{GUP2S}), yields
\begin{equation} \label{GUP22S}
(\Delta x)_{\text{min}}= \sqrt{ 3\sqrt{\pi/2} q\bar\beta/4}.
\end{equation}
It is worth noticing that both large and small droplet-GUP of Eqs.~(\ref{GUP2L}) and (\ref{GUP2S}) are valid only for small $\bar\beta$.
Clearly, for $\bar\beta=0$ one recovers the standard HUP. 
Equations (\ref{GUP22L}) and (\ref{GUP22S}) reveal the existence of modified minimum measurable lengths associated with large and small droplet-GUP.

Figure \ref{fig2} shows that the predicted large droplet-GUP  limits the minimum uncertainty in atoms position. For instance, for $\bar\beta=0.5$, and $m=4$, 
the resulting minimum length is $(\Delta x)_{\text{min}} \approx 8.5$ which is comparable with the droplet equilibrium width, $q_0=8.25$. This reflects the accuracy of our method. 
However, the small droplet-GUP has no minimal length and  behaves in a manner resembling the standard HUP for $\bar\beta \leq 0.5$.
On can infer that small droplet-GUP aquires a minimal length only for larger $\bar\beta$. 
In such a situation our perturbative method with super-Gaussian anstatz is no longer valid.


\section{Numerical simulation}

In this section, we numerically solve the GGPE, using the split-step Fourier transform method by
considering the dynamics of an initially super-Gaussian wave packet (\ref{TDvar}).
The primary focus is to compute  the squared momentum distribution obtained from the Fourier transform of the spacial wavefunction, the density profiles, 
and droplet width for several values of the dimentionless deformation parameter $\bar{\beta}$.

\subsection{ Density profiles}

\begin{figure}[htpb]
    \centering
    \includegraphics[scale=0.5]{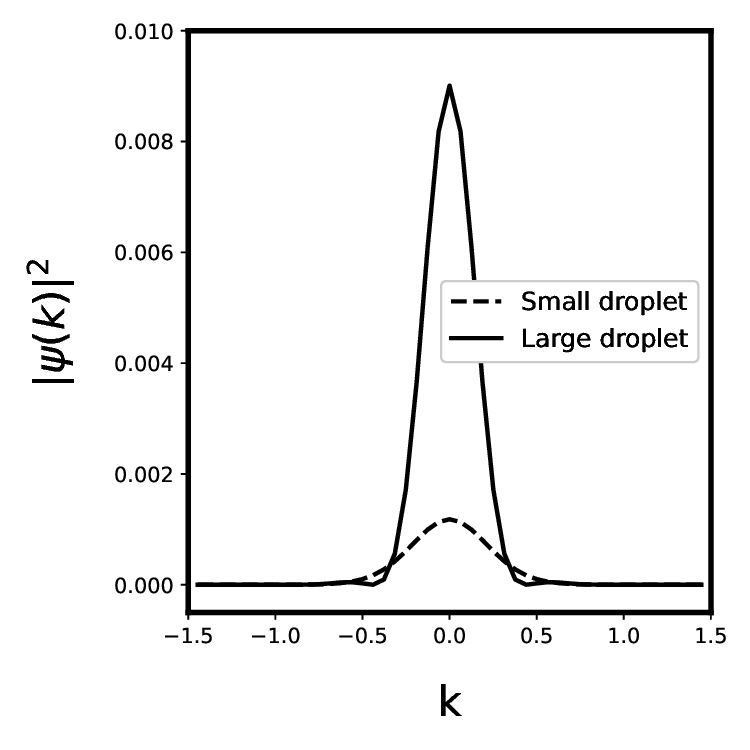}
    \caption{ Droplet density in Fourier space, $|\psi(k)|^2$, with $\bar{\beta}=0.001$,  for both small and large self-bound droplets.}
\label{Four}
\end{figure}

In Fig.~\ref{Four} we present the Fourier transform of profiles $|\psi(k)|^2$ for both small and large self-bound droplets.
We see that $|\psi(k)|^2$ of the large droplet characterized by a uniform flat-top shape develops lateral dips probably due to the interplay of QG and phonon modes which 
propagate through the flat-density bulk and reflected by edges of the droplet \cite{Petrov2}.
However, such a deformation disappears in the case of a small droplet (see dashed line) since the QG action is not strong enough to dominate the quantum fluctuations and the kinetic term.
Here the droplet  can be emitted as small amplitude waves compared to large quantum droplets.

\begin{figure}[h]
    \centering
    \includegraphics[scale=0.38]{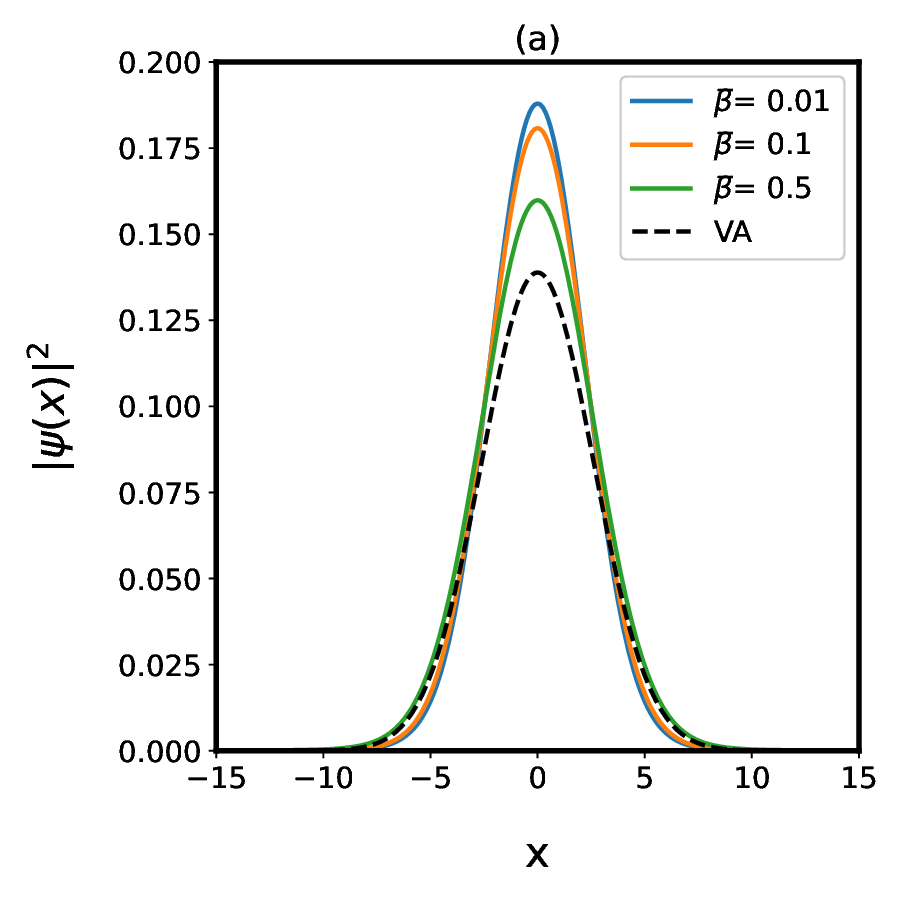}
    \includegraphics[scale=0.38]{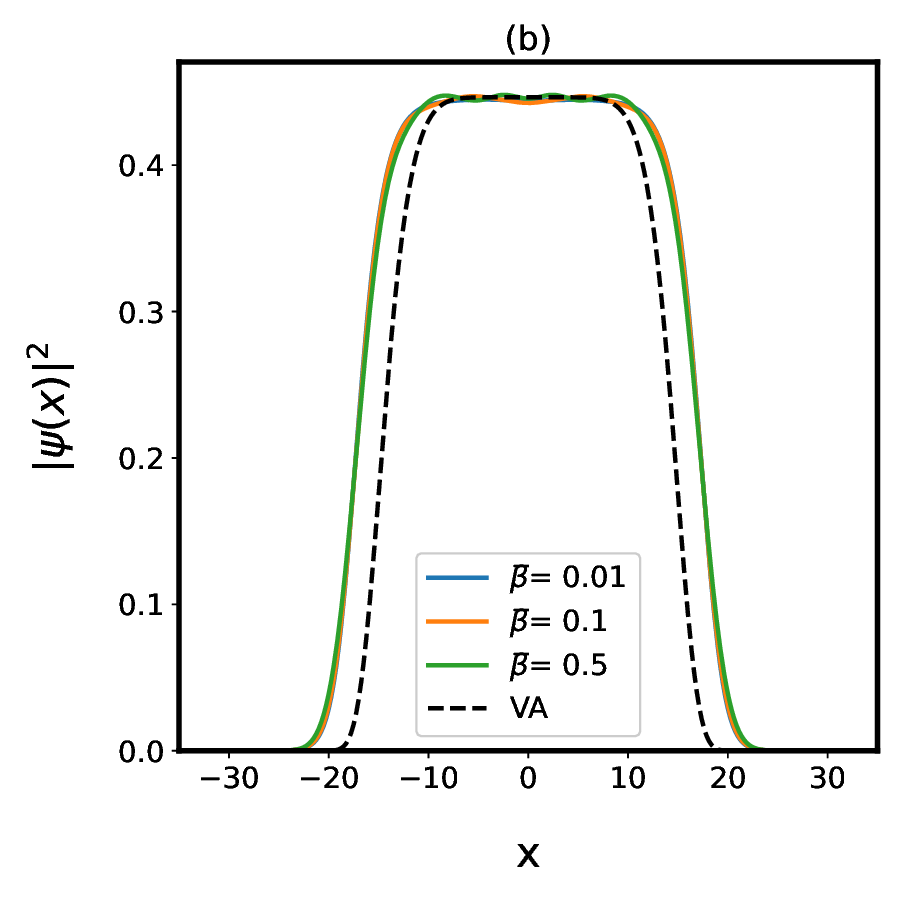}
    \caption{ Density profiles $|\psi(x)|^2$ for different values of $\bar{\beta}$, (a) corresponding to a small Gaussian-like droplet
and (b) to a large flat-top droplet. Dashed lines represent the variational solutions.}
    \label{reaLS}
\end{figure}

Figure ~\ref{reaLS} illustrates effects of the GUP parameter, $\bar \beta$, on the density profiles as predicted by variational calculation and by numerical simulations 
based on the GGPE (\ref{GGPE-GUP1}).
As $\bar \beta$ gets increased, the spatial size of the small self-bound droplet increases  while its amplitude decreases (see Fig.~\ref{reaLS} (a)).
This indicates that the QG acts as an extra force that dilates the droplet.
Our numerical simulations reveal on the other hand that the density of the large droplet exhibits periodic oscillations in the plateau (flat-top) region when $\bar \beta$ becomes stronger
(see Fig.~\ref{reaLS} (b)).  Nevertheless, the QG parameter has no practical significance  close to the edge of the droplet.
Additionally, as is evident from Fig.~\ref{reaLS}, the ground-state density profiles of both small and large droplets are well predicted by the super-Gaussian ansatz, 
affording a perfect modeling approach.


\subsection{Droplet width}

\begin{figure}[h]
	\centering
	\includegraphics[scale=0.5]{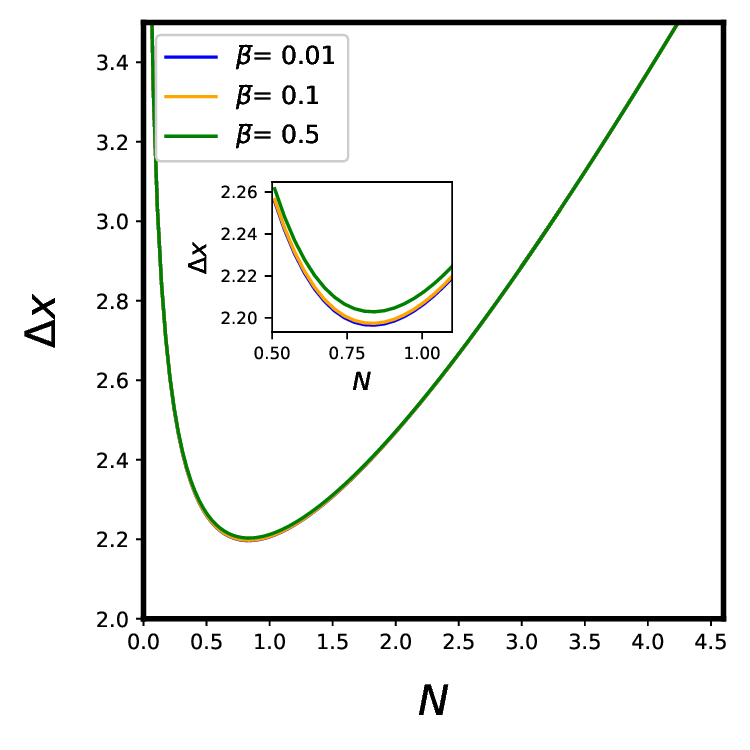}
	\caption{The droplet width as a function of the norm, $N$, for different values of $\bar{\beta}$.}
	\label{size}
\end{figure}

In Fig.~\ref{size} we report the droplet width, $\Delta x$, from Eq.~(\ref{Uncer}) versus the norm, $N$, for  different values of $\bar{\beta}$.
We observe that the droplet size  initially decreases for $N<1$, it reaches its minimal value at $N \sim 1$ (i.e. close to the critical atom number),
then it increases linearly for large $N$. This behavior holds true for any values of $\bar{\beta}$.
Another important remark is that the QG parameter leads to slightly reduce the depth of the local minimum.

\section{Experimental test}

In order to check our theoretical predictions, we compute bounds on the GUP parameter, $\beta_0$, and the corresponding minimal length, associated with our obtained quadratic GUP model, 
using recent experimental data for ${}^{39}$K quantum droplets \cite{Sem}.

\begin{table}[h!]
\begin{center}
\begin{tabular} { cccc cc} 
 \hline \hline\\
                          & $a$  & $a_{12}$   & $\beta_0$ & $(\Delta x)_{\text{min}}$ \\ 
 \hline\\
Large droplet  \,     & 60.9 $a_0$\,    & -53.5 $a_0$\,    & $~ 10^{49}$ \,  &  $\sim 0.8$\,nm \\   
 \hline \\
Small droplet \,  & 60.9 $a_0$    & -53.5 $a_0$    &  $~10^{53}$  &  $\sim 21.9$\,nm \\ \\
 \hline\hline
\end{tabular}
\end{center}
\caption{Upper bounds on the GUP parameters, $\beta_0$, and minimal length, $(\Delta x)_{\text{min}}$, obtained from Eqs.~(\ref{GUP22L}) and (\ref{GUP22S}) 
for ${}^{39}$K quantum droplets \cite{Sem}.}
\label{table:1}
\end{table}

Table \ref{table:1} shows that our model predicts the GUP parameters approximately $\beta_0 \approx 10^{49}$, which although not the better found in the literature, but improves GUP parameters 
resulting from various effects such as corrections in Landau levels  $10^{50}$, quantum noise $10^{57}$ \cite{Bosso},  optics $10^{55}$ \cite{Bra}, and weakly-interacting BEC $10^{56}$ \cite{Boudj}.
We see also that our model predicts a minimum measurable length $(\Delta x)_{\text{min}}$ around $0.8$\,nm for large droplets which is smaller by two orders than
the experimental results for ultracold neutron energy levels in a gravitational quantum well $(\Delta x)_{\text{min}}= 2.41$\,nm \cite{Brau,Nesv} and than that
predicted for graphene $(\Delta x)_{\text{min}} \sim 2.3$ nm \cite{Menc}.
For small droplets, we find $(\Delta x)_{\text{min}}= 21.9$\,nm which is close to that obtained from the Casimir effect for the minimal distance, 29 -58 nm \cite{Frass}.



\section{Conclusions} \label{concl}

We studied the possibility of probing quantum gravity effects in an utradilute self-bound quantum droplet of 1D symmetric Bose mixtures.
We constructed a GGPE-GUP relation which inherently takes into account gravitational corrections using variational and numerical means for both small and large droplets.
The solutions show the existence of a minimal length depending on the system parameters. 
We demonstrated that QG corrections may significantly modify the binding energy, the density profiles, and the equilibrium size of the self-bound quantum droplets.
We discussed also how droplets made of a $^{39}$K Bose mixture might be used to provide upper bounds on the GUP parameter and on the minimal length through a comparison
with exisiting theoretical results and experimental data.
Our findings suggest considerably improved bounds on the GUP parameter, and on the corresponding minimal length.

It is worth noticing that our model can be stratforwardally extended to higher dimensions.
We argue that self-bound droplets may offer a profound opportunity to illuminate the path towards establishing a connection between microscopic and macroscopic worlds.









\begin{thebibliography}{28}


\bibitem{penrose2014} R. Penrose, On the gravitization of quantum mechanics 1: Quantum state reduction, Found. Phys. {\bf 44}, 557 (2014). 
\bibitem{Amati} D. Amati, M. Ciafaloni, G. Veneziano, Phys. Lett. B {\bf 197}, 81 (1987).
\bibitem {Penrose1} R. Penrose, Gen. Relativ. Gravit. {\bf 28}, 581 (1996).
\bibitem {Brons} M. Bronstein Gen. Relativand Gravit. {\bf 44}, 267,(2012).
\bibitem{Bose} S. Bose, A. Mazumdar, G. W. Morley, H. Ulbricht, M. Toros, M. Paternostro, A. A. Geraci, P. F. Barker, M. S. Kim, and G. Milburn, Phys. Rev. Lett. {\bf 119}, 240401 (2017).
\bibitem{Belen} A. Belenchia, R. M. Wald, F. Giacomini, E. Castro-Ruiz, C. Brukner, and M. Aspelmeyer, Phys.Rev.D {\bf 98}, 126009 (2018).
\bibitem{Howl}  R. Howl, V. Vedral, D. Naik, M. Christodoulou, C. Rovelli and A. Iyer, Phys. Rev. X Quantum, {\bf 2}, 010325 (2021).
\bibitem{Piko}  I. Pikovski, M;R. Vanner,  M. Aspelmeyer, M. Kim, C. Brukner,  Nat. Phys. {\bf 8}, 393 (2012). 
\bibitem{Heisenberg} W. Heisenberg, Z. Phys. {\bf 43}, 172 (1927).
\bibitem{Ali} A. F. Ali, S. Das, E. C. Vagenas, Phys.Rev.D {\bf 84}, 044013 (2011).
\bibitem{Scard1} F. Scardigli, Phys. Lett. B {\bf 452}, 39 (1999). 
\bibitem{Mag} M. Maggiore, Phys. Lett. B {\bf 304}, 65 (1993);  Phys. Lett. B {\bf 319}, 83 (1993); Phys. Rev. D {\bf 49}, 5182 (1994).
\bibitem{Kempf}  A. Kempf, G. Mangano and R. B. Mann, Phys. Rev. D {\bf 52}, 1108 (1995).
\bibitem{Scar} F. Scardigli, Phys. Lett. B {\bf 452}, 39 (1999).
\bibitem{Casa}  R.Casadio and F.Scardigli, Phys. Lett. B {\bf 807}, 135558 (2020).
\bibitem{Brack} T. Brack, B. Zybach, F. Balabdaoui, S. Kaufmann, F. Palmegiano, J.-C. Tomasina, S. Blunier, D. Scheiwiller, J. Fankhauser, J. Dual, Nat. Phys. {\bf 18}, 952–957 (2022).
\bibitem{Scard} F. Scardigli, R. Casadio, Eur. Phys. J. C {\bf 75}, 425 (2015).
\bibitem{Bishop} M. Bishop, J. Contreras, and D. Singleton, Universe, {\bf 8}, 192 (2022).
\bibitem{Gao} D. Gao, M. Zhan, Phys. Rev. A {\bf 94}, 013607 (2016).
\bibitem{Fuchs} Fuchs et al., Sci. Adv. {\bf 10}, eadk  2949 (2024).
\bibitem{Simon} S. A. Haine, New J. Phys., {\bf 23}, 033020, (2021).

\bibitem{Shir} K. Shiraishi, Prog. Theor. Phys. {\bf 77} 975  (1987).
\bibitem{Bris} F. Briscese, M. Grether and M. de Llano, Euro. Phys. Lett.,  {\bf 98}, 6 (2012). 
\bibitem{Bris1} F. Briscese, Phys. Lett. B, {\bf 718},  214 (2012).
\bibitem{Dos} M. M. Dos Santos, T. Oniga, A. S. Mcleman, M. Caldwell and C. H. - T. Wang,  J. Plasma Physics , {\bf 79}, 437–442 (2013).
\bibitem{Hans} J. Hansson, S. Francois, International Journal of Modern Physics D, {\bf 26}, 1743003 (2017).
\bibitem{Jaf} M. Jaffe, P. Haslinger, V. Xu, P. Hamilton, A. Upadhye, B. Elder, J. Khoury and H. M\"uller, Nat. Phys {\bf 13},  938 (2017).
\bibitem{Das} S. Das and M. Fridman, Phys. Rev. D {\bf 104}, 026014 (2021). 
\bibitem{Boudj} A. Boudjem\^{a}a, Eur. Phys. J. Plus {\bf 137}, 256  (2022). 
\bibitem{Boudj1} A Tahar Taiba, A. Boudjem\^{a}a, Phys. Lett. B {\bf 861}, 139291 (2025).


\bibitem{Petrov} D. S. Petrov,  Phys. Rev. Lett. {\bf 115}, 155302 (2015).
\bibitem{Petrov1} D. S. Petrov and G. E. Astrakharchik,  Phys. Rev. Lett. {\bf 117}, 100401 (2016).
\bibitem{Pfau} H. Kadau, M. Schmitt, M. Wenzel, C.Wink, T. Maier, I.Ferrier-Barbut, and T. Pfau, Nature {\bf 530}, 194 (2016).
\bibitem{Pfau1}  M. Schmitt, M. Wenzel, F. B\"ottcher, I. Ferrier-Barbut, and T. Pfau, Nature (London) {\bf 539}, 259 (2016).
\bibitem{Chom} L. Chomaz, S. Baier, D. Petter, M. J. Mark, F. W\"achtler, L. Santos and F. Ferlaino, Phys. Rev. X {\bf 6}, 041039 (2016). 
\bibitem{Cab}  C. R. Cabrera, L. Tanzi, J. Sanz, B. Naylor, P. Thomas, P. Cheiney, L. Tarruell,  Science {\bf 359}, 301 (2018);
                       P. Cheiney, C. R. Cabrera, J. Sanz, B. Naylor, L. Tanzi, and L. Tarruell, Phys. Rev. Lett. {\bf 120}, 135301 (2018).
\bibitem{Sem} G. Semeghini, G. Ferioli, L. Masi, C. Mazzinghi, L. Wolswijk, F.Minardi, M. Modugno, G. Modugno, M. Inguscio, and M. Fattori,  Phys. Rev. Lett. {\bf 120}, 235301 (2018).
\bibitem{LHY} T. D. Lee, K. Huang and C. N. Yang, Phys. Rev {\bf 106}, 1135 (1957).
\bibitem{Bors} V. Borsevici, S. Ganguly and G. Manna, arXiv:2411.11047v3 (2025).

\bibitem{Jack} R. Jackiw, S.-Y. Pi, Soliton Solutions to the Gauged Nonlinear Schrödinger Equation on the Plane, Phys. Rev. Lett. {\bf 64}, 2969 (1990). 
\bibitem{Rud} L. Rudnicki, J. Phys. A: Math. Theor. {\bf 49}, 375301 (2016).
\bibitem{Bra} M. C. Braidotti, Z. H. Musslimani, C. Conti, Physica D {\bf 338}, 34 (2017).
\bibitem{Budi} A. Budiyono and H. K. Dipojono Phys. Rev. A {\bf 102}, 012205 (2020).

\bibitem{Fit} T. Fityo,  Phys. Lett. A {\bf 372}, 5872 (2008).
\bibitem{VaK} B. Vakili, M. A. Gorji,  J. Stat. Mech. P10013 (2012).
\bibitem{Cast} E. Castellanos and C. Laemmerzahl, Phys. Lett. B {\bf 731}, 1 (2014).
\bibitem{Zhang}  X. Zhang and C. Tian,  Chinese. Phys. Lett. {\bf 32}, 010303 (2015).
\bibitem{Li}  H. L. Li,  J. X. Ren, W. W. Wang, B. Yang and H. J. Shen, J. Stat. Mech. 023106 (2018).
\bibitem{Sanj} S. Dey, V. Hussin,  International Journal of Theoretical Physics {\bf 58}, 3138 (2019). 

\bibitem{Boudj3} K. Mohammed Elhadj, L. Al Sakkaf,  A. Boudjem\^{a}a, U. Al Khawaja, Phys. Lett. A {\bf 494}, 129274 (2024).
\bibitem{Das2} S. Das, and E.C. Vagenas,  Phys. Rev. Lett. {\bf 101}, 221301 (2008).
\bibitem{Fadel} M. Fadel, M.Maggiore, Phys. Rev. D {\bf 105}, 106017 (2022).
\bibitem{Matt} Matthew J. Lake, Ukrainian Journal of Physics. { \bf 64}, 1036 (2019).

\bibitem{Karl} M. Karlsson,  Phys. Rev. A {\bf 46}, 2726 (1992) .
\bibitem{Tsy} E. N. Tsoy, A. Ankiewicz, and N. Akhmediev,  Phys. Rev. E {\bf 73}, 036621 (2006).
\bibitem{Baiz}  B. B. Baizakov, A. Bouketir, A. Messikh, A. Benseghir, and B. A. Umarov, Int. J. of Mod. Phys. B {\bf 25}, 2427  (2011).
\bibitem{Petrov2} M. Tylutki, G. E. Astrakharchik, B. A. Malomed, and D. S. Petrov, Phys. Rev. A {\bf 101}, 051601(R) (2020).

\bibitem{Bosso} P. Bosso {\it et. al} Class. Quantum Grav. {\bf 40}, 195014 (2023).
\bibitem{Nesv} V. V. Nesvizhevsky {\it et al}., Nature (London) {\bf 415}, 297 (2002); V. V. Nesvizhevsky {\it et al}., Phys. Rev. D {\bf 67}, 102002 (2003).
\bibitem{Brau} F. Brau and F. Buisseret, Phys. Rev. D {\bf 74}, 036002 (2006).
\bibitem{Menc}  L. Menculini, O. Panella, P. Roy, Phys. Rev. D {\bf 87}, 065017 (2013).
\bibitem{Frass} A. M. Frassino and O. Panella, Phys. Rev. D {\bf 85}, 045030 (2012).


\end{thebibliography}
\end{document}